\documentstyle[12pt]{article}
\begin{document}
\[\]
\[\]
\begin{center}
{\large \bf Dynamics of general relativistic spherically
symmetric dust thick shells}
\[\]
{\bf S. Khakshournia\footnote{Email Address:
skhakshour@seai.neda.net.ir}
and R. Mansouri\footnote{Email Address: mansouri@sharif.edu}} \\

Department of physics, Sharif University of Technology, Tehran, Iran
\\
\end{center}
\[\]
\begin{center}
{\bf \ Abstract}
\end{center}
We consider a spherical thick shell immersed in two different
spherically symmetric space-times. Using the fact that the
boundaries of the thick shell with two embedding space-times must
be nonsingular hypersurfaces, we develop a scheme to obtain the
underlying equation of motion for the thick shell in general. As
a simple example, the equation of motion of a spherical dustlike
shell in vacuum is obtained. To compare our formalism with the
thin shell one, the dynamical equation of motion of the thick
shell is then expanded to the first order of its thickness. It is
easily seen that the thin shell limit of our dynamical equation is
exactly that given in the literature for the dynamics of a thin
shell. It turns out that the effect of thickness is to speed up
the collapse of the shell.
\newpage
\subsection*{I. Introduction}
The thin shell formalism of general relativity has found wide
applications in general relativity and cosmology[1-3]. Studies on
gravitational collapse, dynamics of bubbles and domain walls in
inflationary models, wormholes, signature changes, structure and
dynamics of voids in the large scale structure of the universe
are some of the applications. Thin shells are considered as
idealized zero thickness objects, with a $\delta$-function
singularity in their energy-momentum and Einstein tensors. This
is regarded to be an idealization of a real shell with a finite
thickness. However, the dynamics of a real thick shell has been
rarely discussed in the literature because of the complexity one
is faced with when trying to define it within general relativity
and to find its exact underlying dynamical equations. The
outstanding paper that modifies the Israel thin shell equations
to treat the motion of spherical and planar thick domain walls is
that of Garfinkle and Gregory[4]. Their work deals with an
expansion of the coupled Einstein-scalar equations in powers of
the thickness of the domain wall (see also [5]). According to the
results of that paper, the effect of thickness in the first
approximation is to reduce effectively the energy density of the
wall compared to the corresponding thin domain wall, and
therefore to increase the
collapse velocity of the wall in vacuum.\\
 In this paper we first generally suggest the proper matching
conditions on the boundaries of the spherical thick shell
embedded in an inner and an outer spherically symmetric
space-time. As a simple example, these matching conditions are
then used to investigate the motion of spherical dustlike thick
shell in vacuum.
\subsection*{II. The junction conditions}
Consider a spherically symmetric thick shell with two boundaries
$\Sigma_{1}$ and $\Sigma_{2}$ dividing the space-time into
three-regions: ${\cal M}_{in}$ for inside the inner boundary
$\Sigma_{1}, {\cal M}_{out}$ for outside the outer boundary
$\Sigma_{2}$, and ${\cal M}$  for the thick shell having two
boundaries $\Sigma_{1}$ and $\Sigma_{2}$. First of all, let us
write down the appropriate junction condition on each boundary
$\Sigma_{j}$ (j=1,2) treated as a (2+1)-dimensional timelike
hypersurface. We expect the continuity of the second fundamental
form of $\Sigma_{j}$, or the extrinsic curvature tensor $K_{ab}$
 of $\Sigma_{j}$, so that we can consider $\Sigma_{1} (\Sigma_{2})$ as a
 boundary surface separating ${\cal M}$ region from ${\cal M}_{in}$
 (${\cal M}_{out}$). This crucial requirement is formulated as
\begin{equation}
\left[ K_{ab} \right] \stackrel{\Sigma_{j}}{=} 0 \quad\quad\quad\quad
 \quad \quad (j=1,2),
\end{equation}
where the square bracket indicates the jump of $K_{ab}$ across
$\Sigma_{j}$, Latin indices range over the intrinsic coordinates
of $\Sigma_{j}$ denoted by $(\tau_{j} , \theta , \varphi )$,
where $\tau_{j}$ is the proper time of $\Sigma_{j}$. In
particular, the angular component of Eq. (1) on each boundary is
written as
\begin{eqnarray}
K_{\theta}^{\theta^{+}} \Bigl|_{\Sigma_{1}} -K_{\theta}^{\theta^{-}}
\Bigr|_{\Sigma_{1}}=0 , \\
K_{\theta}^{\theta^{+}} \Bigl|_{\Sigma_{2}} -K_{\theta}^{\theta^{-}}
\Bigr|_{\Sigma_{2}}=0 ,
\end{eqnarray}
where the superscript $+(-)$ refers to the side of $\Sigma_{j}$
towards which the corresponding unit spacelike normal vector
$n^{\alpha} (-n^{\alpha})$ points. This means that on $\Sigma_{1}
(\Sigma_{2})$, the superscript + refers to the region ${\cal M}
({\cal M}_{out})$ and the superscript $-$ refers to the region
${\cal M}_{in}(\cal M )$. Adding Eqs. (2) and (3), we obtain
equation
\begin{equation}
K_{\theta}^{\theta^{+}} \Bigl|_{\Sigma_{2}}
-K_{\theta}^{\theta^{-}}
\Bigl|_{\Sigma_{1}}+K_{\theta}^{\theta^{+}} \Bigl|_{\Sigma_{1}}-
K_{\theta}^{\theta^{-}} \Bigl|_{\Sigma_{2}}=0.
\end{equation}
In the next section we will apply the general equation (4) to the
special case of a collapsing spherical dust shell in vacuum.
\subsection*{III. Collapse of a spherical dust thick shell in vacuum}
Consider the Lemaitre-Tolman-Bondi (LTB) metric to describe the
dust thick shell. In the synchronized comoving coordinates $(\tau
, r, \theta , \varphi )$ the metric is written in the form [6]
\begin{equation}
ds^{2}=-d\tau^{2} +\frac{R'^{2}}{1+E(r)} dr^{2} +R^{2} (r, \tau ) (d\theta^{2}
+\sin^{2} \theta d\varphi^{2}),
\end{equation}
where overdot and prime denote partial differentiation with respect to $\tau$
and $r$, respectively, and $E(r)$ is an arbitrary real function such that
$E(r)>-1$. Then the corresponding Einstein field equations turn out to be
\begin{eqnarray}
\dot{R}^{2} (r, \tau ) =E(r) +\frac{F(r)}{R}, \\
8\pi G\rho (r, \tau ) =\frac{F'(r)}{R^{2}R'},
\end{eqnarray}
where $\rho (r, \tau )$ is the energy density of the matter fluid
in ${\cal M}$, and $F(r)$ is another arbitrary real smooth
function such that $F(r)>0$. Furthermore, in order to avoid shell
crossing of dust matter during their radial motion, we require
$R'(r, \tau )>0$. This together with the assumption of positive
mass density $\rho (r, \tau )>0$, implies that $F'(r)\geq 0$. The
 induced intrinsic metric on $\Sigma_{j}$ may be represented as
\begin{equation}
ds^{2}\Bigl|_{\Sigma_{j}} =-d\tau^{2}_{j} +R^{2}_{j} (\tau_{j}) (d\theta^{2} +
\sin^{2}\theta d\varphi^{2}) \quad\quad\quad\quad (j=1,2),
\end{equation}
where $R_{j}(\tau_{j})$ being the proper radius of $\Sigma_{j}$.
For simplicity, we may assume that the boundaries $\Sigma_{1}$
and $\Sigma_{2}$ are comoving with respect to the LTB geometry.
This requires that the peculiar velocity of $\Sigma_{j}$ measured
by the comoving observers of ${\cal M}$ to be zero. Then the
matching relations yield
 \begin{equation}
 \tau =\tau_{j} +{\rm constant} \quad \quad \quad \quad \quad \quad \quad(i=1,2).
 \end{equation}
 Now, we may define the constant comoving thickness of the shell as follows
\begin{equation}
2\delta =r_{2}-r_{1},
\end{equation}
where $r_{1}$ and $r_{2}$ are comoving radii of the boundaries
$\Sigma_{1}$ and $\Sigma_{2}$, respectively. We assume now the
spherical dust thick shell to be immersed in vacuum. In this
case, the space-time exterior to the shell is Schwarzschild, and
the interior is taken to be Minkowski flat space-time. Different
terms appearing in the equation (4) may now be explicitly
derived. Using the metric (5) together with Eqs. (6), (7) and
(9), we compute the relevant extrinsic curvature tensors in the
region $\cal M$ as
\begin{equation}
K_{\theta}^{\theta^{+}} \Bigl|_{\Sigma_{1}} =\frac{1}{R_{1}}
\sqrt{1+\dot{R}_{1}^{2}-\frac{F(r_{1})}{R_{1}}} \ \ , \ \
K_{\theta}^{\theta^{-}} \Bigl|_{\Sigma_{2}} =\frac{1}{R_{2}}
\sqrt{1+\dot{R}_{2}^{2}-\frac{F(r_{2})}{R_{2}}} \ \ ,
\end{equation}
where $R_{j}\equiv R(r_{j},\tau )$. Furthermore, the following
expressions for the relevant extrinsic curvature tensors in
${\cal M}_{in}$ and ${\cal M}_{out}$ also hold [3]
\begin{equation}
K_{\theta}^{\theta^{-}} \Bigl|_{\Sigma_{1}} =\frac{1}{R_{1}}
\sqrt{1+\dot{R}_{1}^{2}} \ \ , \ \ K_{\theta}^{\theta^{+}}
\Bigl|_{\Sigma_{2}} =\frac{1}{R_{2}}
\sqrt{1+\dot{R}_{2}^{2}-\frac{{\cal R} (r_{2})}{R_{2}}} \ \ ,
\end{equation}
where ${\cal R} (r_{2})$ is the Schwarzschild radius of the
spherical shell within the comoving surface $r_{2}$. Now, to
obtain the dynamical equation of the thick shell, we first expand
the following quantities in a Taylor series around $r_{0}$, the
mean comoving radius of the thick shell:
\begin{eqnarray}
R(r_{j},\tau )=R(r_{0},\tau )+\epsilon_{j} \delta R'(r_{0}, \tau
)+ {\cal O} (\delta^{2}),
\end{eqnarray}
\begin{eqnarray}
F(r_{j})=F(R_{0})+\epsilon_{j} \delta F'(r_{0}) +{\cal O}
(\delta^{2}),
\end{eqnarray}
\begin{eqnarray}
 {\cal R} (r_{2}) ={\cal R}(r_{0})+
\delta {\cal R'} (r_{0}) + {\cal O} (\delta^{2}),
\end{eqnarray}
where $\epsilon_{1}=-1$ and $\epsilon_{2}=+1$. Using Eqs. (13),
(14) and (15) in the expressions (11) and (12) and keeping only
terms up to the first order of $\delta$, we obtain
\begin{eqnarray}
K_{\theta}^{\theta^{-}} \Bigl|_{\Sigma_{1}} =\frac{1}{R_{0}}
\sqrt{1+\dot{R}_{0}^{2}} \left( 1+\delta \Bigl(
\frac{R'_{0}}{R_{0}} -\frac{\dot{R}_{0}
\dot{R}'_{0}}{1+\dot{R}_{0}^{2}} \Bigr) \right)\ ,
\end{eqnarray}
\begin{eqnarray*}
&&K_{\theta}^{\theta^{+}} \Bigl|_{\Sigma_{2}} =\frac{1}{R_{0}}
\sqrt{1+\dot{R}_{0}^{2} -\frac{{\cal R} (r_{0})}{R_{0}}} \left(
1-\delta \Bigl( \frac{R'_{0}}{R_{0}} -\frac{\dot{R}_{0}
\dot{R'}_{0} -{{\cal R'}(r_{0}) \over 2R_{0}} +{R'_{0} {\cal R}
(r_{0})\over 2R_{0}^{2}}}{1+\dot{R}_{0}^{2}
-{{\cal R} (r_{0})\over R_{0}}}\Bigr) \right)\ , \\
&&K_{\theta}^{\theta^{+}} \Bigl|_{\Sigma_{1}} =\frac{1}{R_{0}}
\sqrt{1+\dot{R}_{0}^{2} -\frac{F(r_{0})}{R_{0}}} \left( 1+\delta
\Bigl( \frac{R'_{0}}{R_{0}} -\frac{\dot{R}_{0} \dot{R'}_{0} -{F'(r_{0})
\over 2R_{0}} +{R'_{0} F (r_{0})\over 2R_{0}^{2}}}{1+\dot{R}_{0}^{2}
-{F (r_{0})\over R_{0}}}\Bigr) \right)\ , \\
&&K_{\theta}^{\theta^{-}} \Bigl|_{\Sigma_{2}} =\frac{1}{R_{0}}
\sqrt{1+\dot{R}_{0}^{2} -\frac{F(r_{0})}{R_{0}}} \left( 1-\delta
\Bigl( \frac{R'_{0}}{R_{0}} -\frac{\dot{R}_{0} \dot{R'}_{0}
-{F'(r_{0}) \over 2R_{0}} +{R'_{0} F (r_{0})\over
2R_{0}^{2}}}{1+\dot{R}_{0}^{2} -{F(r_{0})\over R_{0}}}\Bigr)
\right)\ ,
\end{eqnarray*}
where $R_{0}\equiv R(r_{0},\tau )$. Substituting Eq. (16) into
Eq. (4) and noting that for the metric LTB, $F(r_{0})$ plays just
the role of the Schwarzschild radius of the part of the spherical
thick shell within the comoving surface $r_{0}$ denoted by ${\cal
R}(r_{0})$, we obtain after some rearrangement the dust thick
shell's equation
of motion written up to the first-order in $\delta$:\\[0.5cm]
$\alpha - \beta = 2\delta \frac{F'(r_{0})}{2R_{0}
\sqrt{1+\dot{R}_{0}^{2} - {F(r_{0})\over R_{0}}}}$
\begin{eqnarray}
-\delta \left(\frac{R'_{0}}{R_{0}} (\alpha -\beta )+\dot{R}
\dot{R'}_{0} \Bigl( \frac{\alpha - \beta}{\alpha \beta} \Bigr)
+\frac{1}{2\beta R_{0}} \Bigl( {\cal R'} (r_{0})
+\frac{R'_{0}}{R_{0}} {\cal R} (r_{0})\Bigr)\right),
\end{eqnarray}
with
\begin{equation}
\alpha \equiv \sqrt{1+\dot{R}_{0}^{2}} \quad , \quad \beta
=\sqrt{1+\dot{R}_{0}^{2} -\frac{{\cal R}(r_{0})}{R_{0}}}.
\end{equation}
This is the generalization of thin shell dynamical equation up to
the first order of the thickness. It is now interesting to verify
the thin shell limit of this thick shell dynamical equation. In
order to do this, consider the following definition for the
surface energy density of the infinitely thin shell[1]:
\begin{equation}
\sigma =\int_{-\epsilon}^{\epsilon} \rho (r, \tau ) dn,
\end{equation}
where $n$ is the proper distance in the direction of the normal
$n_{\mu}$ and $2\epsilon$ is the physical thickness of the shell.
With the metric (5), Eq. (19) takes the form
\begin{equation}
\sigma =\int_{-\delta}^{\delta}\rho(r,\tau)\frac{R'(r,\tau)}{\sqrt
{1+E(r)}}dr.
\end{equation}
Using Eqs. (6) and (7), we find that Eq. (20) can be written as
\begin{equation}
8\pi G\sigma =\int_{-\delta}^{\delta} \frac{F'(r)}{R^{2}
\sqrt{1+\dot{R}^{2} -{F(r)\over R}}} dr  \cdot
\end{equation}
Using Eqs. (13) and (14), we may integrate Eq. (21) up to the
first order in $\delta$ to get
\begin{equation}
8\pi G\sigma =2\delta\frac{F'(r_{0})}{R_{0}^{2}
\sqrt{1+\dot{R}_{0}^{2} - {F(r_{0})\over R_{0}}}}+{\cal O}
(\delta^{2}).
\end{equation}
Substituting (22) into Eq. (17), we get the following result\\[0.5cm]
$\alpha - \beta =4\pi G\sigma R_{0}$
\begin{equation}
 -\delta \left(
\frac{R'_{0}}{R_{0}} (\alpha - \beta ) +\dot{R}_{0} \dot{R'}_{0}
(\frac{\alpha - \beta}{\alpha \beta})+\frac{1}{2\beta R_{0}}
({\cal R'} (r_{0})+\frac{R'_{0}}{R_{0}} {\cal R} (r_{0}))\right).
\end{equation}
Note that in the zero thickness limit of the shell, as $\delta
\longrightarrow 0$, the second term on the right hand side of Eq.
(23) is regular and goes to zero such that Eq. (23) reduces to
the Israel's equation of motion for the dust thin shell in vacuum,
as given by Israel [1], Ipser-Sikivie[7], Sato[8], and Khorrami-Mansouri [9].\\
 To see the explicit effect of the thickness on the dynamics of the thick shell
we rewrite Eq. (23) as
 \begin{equation}
 \alpha - \beta = 4\pi G\tilde{\sigma} R_{0},
 \end{equation}
 where $\tilde{\sigma}$, the effective surface density, is defined by
\begin{equation}
\tilde{\sigma} =\sigma -\frac{\delta}{4\pi GR_{0}} \left(
\frac{R'_{0}}{R_{0}} (\alpha - \beta ) +\dot{R}_{0} \dot{R'}_{0}
\Bigl( \frac{\alpha - \beta}{\alpha \beta} \Bigr)
+\frac{1}{2\beta R_{0}} \Bigl( {\cal R'} (r_{0})
+\frac{R'_{0}}{R_{0}} {\cal R} (r_{0})\Bigr) \right).
\end{equation}
We see that Eq. (24) has the same form as the well-known Israel's
equation for a thin shell with the effective surface density
$\tilde{\sigma}$. Now let us have a closer look at the terms
within the brackets. Note that for a dust shell starting its
collapse at rest, the velocity $\dot{R}_{0}$ is negative during
the collapse, it also becomes more negative with $r$ so that
$\dot{R'}_{0}< 0$, so the combination of
$\dot{R}_{0}\dot{R'}_{0}$ must be positive. On the other hand,
the Schwarzschild radius of the shell layers is increased with
$r$ so that ${\cal R'} (r_{0})>0$. Therefore all the terms within
the bracket on the right hand side of Eq. (25) are positive. This
leads to the result $\tilde{\sigma}<\sigma$.
 Now, solving Eq. (24) for $\dot{R}^{2}$ we find
\begin{equation}
\dot{R}_{0}^{2} =4\pi^{2} G^{2} R_{0}^{2} \tilde{\sigma}^{2}
+\frac{{\cal R}^{2} (r_{0})}{64\pi^{2}G^{2}R_{0}^{4}
\tilde{\sigma}^{2}} +\frac{{\cal R} (r_{0})}{2R_{0}} -1.
\end{equation}
From Eq. (26), it follows that for a given $R_{0}$ and ${\cal R}
(r_{0})$ as long as $R_{0}>{\cal R}(r_{0})$, smaller
$\tilde{\sigma} (\tilde{\sigma} <\sigma )$ leads to a larger
$\dot{R}^{2}$. Therefore the first-order thickness corrections to
the Israel thin shell approximation speed up the collapse of dust
shell in vacuum.
 \subsection*{IV. Conclusion}
 We have presented a simple procedure to investigate the dynamics
of a spherical thick shell embedded in an otherwise spherically
symmetric space-time, based on the Darmois conditions satisfying
on the shell boundaries. As the simplest nontrivial example, we
applied our scheme to the case of the collapse of a thick shell of
dust in vacuum and obtained the zero thickness limit of our
formalism which is just the Israel thin shell equation. It has
been shown that the effect of thickness, up to the first order in
the shell thickness, is to speed up its collapse in vacuum.

\end{document}